\documentclass[12pt]{article}

\usepackage{latexsym}
\usepackage{amsfonts}
\usepackage[usenames,dvipsnames]{xcolor}
\usepackage{graphicx,amssymb,amsmath,epsfig}
\usepackage[english]{babel}
\usepackage{graphicx}% Include figure files
\usepackage{dcolumn}% Align table columns on decimal point
\usepackage{bm}
\usepackage{slashed}
\usepackage{caption}
\definecolor{myblue}{rgb}{0,0,0.8}
\usepackage[colorlinks=true
,urlcolor=myblue	
,anchorcolor=myblue
,citecolor=myblue
,filecolor=myblue
,linkcolor=myblue
,menucolor=myblue
,pagecolor=myblue
,linktocpage=true  %%%%%%  Link no numero da pagina em vez do titulo   %%%%%%%%%%
]{hyperref}

\catcode`\@=11
\def\marginnote#1{}

\newcount\hour
\newcount\minute
\newtoks\amorpm
\hour=\time\divide\hour by60 \minute=\time{\multiply\hour by60
\global\advance\minute by-\hour}\edef\standardtime{{\ifnum\hour<12
\global\amorpm={am}%
        \else\global\amorpm={pm}\advance\hour by-12 \fi
        \ifnum\hour=0 \hour=12 \fi
        \number\hour:\ifnum\minute<10
0\fi\number\minute\the\amorpm}}
\edef\militarytime{\number\hour:\ifnum\minute<10 0\fi\number\minute}

\def\draftlabel#1{{\@bsphack\if@filesw {\let\thepage\relax
   \xdef\@gtempa{\write\@auxout{\string
      \newlabel{#1}{{\@currentlabel}{\thepage}}}}}\@gtempa
   \if@nobreak \ifvmode\nobreak\fi\fi\fi\@esphack}
        \gdef\@eqnlabel{#1}}
\def\@eqnlabel{}
\def\@vacuum{}
\def\draftmarginnote#1{\marginpar{\raggedright\scriptsize\tt#1}}
\def\draft{\oddsidemargin -.5truein
        \def\@oddfoot{\sl preliminary draft \hfil
        \rm\thepage\hfil\sl\today\quad\militarytime}
        \let\@evenfoot\@oddfoot \overfullrule 3pt
        \let\label=\draftlabel
        \let\marginnote=\draftmarginnote

\def\@eqnnum{(\theequation)\rlap{\kern\marginparsep\tt\@eqnlabel}%
\global\let\@eqnlabel\@vacuum}  }

\def\numberbysection{\@addtoreset{equation}{section}
        \def\theequation{\thesection.\arabic{equation}}}

\def\underline#1{\relax\ifmmode\@@underline#1\else
 $\@@underline{\hbox{#1}}$\relax\fi}

\catcode`@=12 \relax

\numberbysection
\topmargin by -\headsep
\def\be{\begin{equation}}
\def\ee{\end{equation}}
\def\br{\begin{eqnarray}}
\def\er{\end{eqnarray}}

\def\({\left(}
\def\){\right)}
\def\[{\left[}
\def\]{\right]}

\def\b{\beta}

\def\l{\lambda}

\def\tp0{\Theta_{+}^{(0)}}
\def\tm0{\Theta_{-}^{(0)}}

\def\l{\lambda}

\def\bi{\begin{itemize}}
\def\ei{\end{itemize}}

\begin{document}

\vspace*{1cm}
\noindent

\vskip 1 cm
\begin{center}
{\Large\bf The Lorentz-violating real scalar field at thermal equilibrium}
\end{center}
\normalsize
\vskip 1cm
\begin{center}
{A. R. Aguirre}\footnote{\href{mailto:alexis.roaaguirre@unifei.edu.br}{alexis.roaaguirre@unifei.edu.br}}$^{,\clubsuit}$,
{G. Flores-Hidalgo}\footnote{\href{mailto:gfloreshidalgo@unifei.edu.br}{gfloreshidalgo@unifei.edu.br}}$^{,\clubsuit}$, 
{R. G. Rana}\footnote{\href{mailto:galhardorod@gmail.com}{galhardorod@gmail.com}}$^{,\dagger,\clubsuit}$, 
and 
{E. S. Souza}\footnote{\href{mailto:edsonssouzafis@gmail.com}{edsonssouzafis@gmail.com}}$^{,\ddagger,\clubsuit}$,
\\[.5cm]

\par \vskip .1in \noindent
$^{\clubsuit}$\emph{Instituto de F\'isica e Qu\'imica, Universidade Federal de Itajub\'a, Av. BPS 1303, \\Itajub\'a, MG CEP 37500-903, Brazil.}\\[0.3cm]

$^{\dagger}$\emph{Centro Brasileiro de Pesquisas F\'isicas, Rua Dr. Xavier Sigaud 150, Urca, \\Rio de Janeiro, RJ, Brazil, CEP 22290-180.}\\[0.3cm]

$^{\ddagger}$\emph{Instituto de F\'isica Gleb Wataghin - UNICAMP, 13083-859, Campinas-SP, Brazil.}\\
\vskip 3cm
\end{center}

\begin{abstract}

In this paper we study Lorentz-Violation(LV) effects on the thermodynamics properties of a real scalar field theory due to the presence of a constant background tensor field. In particular, we analyse and compute explicitly  the deviations of the internal energy, pressure, and entropy of the system at thermal equilibrium due to the LV contributions. For the free massless scalar field we obtain exact results, whereas for the massive case we perform approximated calculations. Finally, we consider the self interacting  $\phi^4$ theory, and perform perturbative expansions in the coupling constant for obtaining relevant thermodynamics quantities.  

%\\
\end{abstract}

%\end{titlepage}
\newpage
\tableofcontents
%\maketitle

%\newpage
\vskip .4in

%%%%%%%%%%%%%%%%%%%%%%%%%%%%%%%%%%%%%%%%%%%%%%%%%%%%%%%%%%%%%%%%%

\section{Introduction}
Nowadays, it is well-known that symmetries play a  fundamental role for describing our modern physical theories. In particular, besides gauge symmetries in the standard model of fundamental particle and interactions, the Lorentz and CPT symmetries are the main building blocks of the theory, and are expected to be respected in any physical situation. However,  some years ago, several different theories appeared in the literature, in which apparently such assumption is not longer satisfied \cite{Kos1}--\cite{Kos1.4}, thereby attracting a great attention
in the last few decades both from the theoretical and experimental point of view.

Recently, an interesting extension of the standard model (SME) have been also proposed as an effective field theory, bringing new insights on the possible effects of Lorentz and CPT symmetries breaking \cite{Colla1,Colla2}. The main idea is that these symmetries can be manifestly broken  by incorporating some small constant vector and tensor fields in the theory,  generating in this way preferential directions in spacetime. Then, it is expected that such anisotropies in the spacetime should appear as quite small deviations of any physical measurement predicted by a Lorentz-invariant (LI) theory. See Ref. \cite{Table} (and references therein) to find updated experimental data that constraining many Lorentz-violating (LV) coefficients in the SME. 

The effects of Lorentz violation has been widely investigated in several different scenarios. It is worth mentioning recent studies on  the  statistical mechanics in LV background \cite{Colladay2004}, the Cassimir effect \cite{Cruz2017}--\cite{Erdas}, the Bose-Einstein condensation \cite{Colla2006}--\cite{furtado}, QED sector \cite{Casana2009a}--\cite{Escobar2015}, LV theories with boundary conditions \cite{Borges1,Borges2},  Chern-Simons-like terms \cite{Mariz}--\cite{Brito2009},  scattering processes \cite{Casana2012}--\cite{Khanna}, geometrical correspondences \cite{RiemannFinsler,Colla2019}, supersymmetric LV models \cite{Berger}--\cite{Bernal}, as well as many others interesting subjects  (see also \cite{Kosultimo} and references therein for a more exhaustive list of related papers).

On the other hand, finite temperature effects in field theories have been widely studied  mainly in the context of  cosmological problems \cite{Kirzhnits,Linde1},   symmetry
restoration in theories with spontaneously broken symmetry \cite{Jackiw2,Weinberg74,Linde76}, and more recently in connection with  phase transitions in QCD problems related  with high energy heavy ion collisions \cite{Gross,Anselm1,Anselm2,Bjorken,Raja}. For more articles and textbooks on this subject, see Refs.  \cite{Kapusta,Das,Zinn,Bellac,Mariano}.

In the present work, we are interested in investigating the  finite temperature effects in a the Lorentz-violating real scalar field theory due to the presence of a constant background tensor contribution. The paper is outlined as follows. In next section,  the particle spectrum of the LV model is discussed. In section \ref{free}, we discuss the thermal effects in the LV free scalar theory. The interacting case at finite temperature will be analysed  in section \ref{interacting}. In section \ref{conclusao}, some concluding remarks are presented. Finally, there are some explicit calculations can be found in the appendices.

%%%%%%%%%%%%%%%%%%%%%%%%%%%%%%%%%%%%%%%%%%%%%%%%%%%%%%%%%%%%%%%%%
\section{Particle spectrum}
\label{spectrum}
%%%%%%%%%%%%%%%%%%%%%%%%%%%%%%%%%%%%%%%%%%%%%%%%%%%%%%%%%%%%%%%%%
In this section, we will discuss the spectrum of a theory with a real scalar field in the presence of a  LV background tensor field. 
Let us consider the following lagrangian density for the real scalar $\phi(x)$\cite{Colla1, Colla2},
\begin{equation}
{\cal L}=\frac{1}{2}\partial_\mu\phi(x)\partial^\mu\phi(x)+\frac{\kappa^{\mu\nu}}{2}\partial_\mu\phi(x)\partial_\nu\phi(x)
-\frac{m^2}{2}\phi^2,
\label{phi0}
\end{equation}
where $\kappa_{\mu\nu}$ are dimensionless constant tensor coefficients  for the Lorentz-violation that preserves CPT symmetry \cite{Kos2000, Potting2012}. Since this model is still invariant under space-time translations, the energy and
linear momentum will be conserved. In fact, we get respectively,
\begin{equation}
P^0=H=\frac{1}{2}\int d^3{\bf x} 
\left( \Delta ^{00}\dot{\phi}^2(x)-\Delta^{ij}\partial_i\phi(x)\partial_j\phi(x)+m^2\phi^2(x) \right),
\label{hamiltonian}
\end{equation}
and
\begin{equation}
{\bf P}=-\int d^3 {\bf x}~\!\Delta^{0\mu}\partial_\mu\phi(x)\nabla \phi(x),
\label{momentum}
\end{equation}
where $\Delta^{\mu\nu}=\eta^{\mu\nu}+\kappa^{\mu\nu}$ is a modification of the Minkowski metric $\eta^{\mu\nu}$.
From eq. (\ref{phi0}) we get the field equation,
\begin{equation}
\Delta^{\mu\nu}\partial_\mu\partial_\nu\phi(x)+m^2\phi(x)=0,
\label{eqmot}
\end{equation}
and the canonically conjugate momentum $\pi(x)$, which is given by
\begin{equation}
\pi(x)=\Delta^{0\mu}\partial_\mu\phi(x).
\label{mconjugate}
\end{equation}
In order to quantize the theory, we impose the following equal-time commutation relations,
\begin{equation}
[\hat{\phi}({\bf x},t),\hat{\pi}({\bf y},t)]=i\delta({\bf x}-{\bf y}),\qquad 
[\hat{\phi}({\bf x},t),\hat{\phi}({\bf y},t)]=0,\qquad 
[\hat{\pi}({\bf x},t),\hat{\pi}({\bf y},t)]=0,
\label{commutation}
\end{equation}
where the field operators satisfy equations (\ref{eqmot}) and (\ref{mconjugate}). In this way, to solve the field equation for $\hat{\phi}(x)$, we expand in plane waves as follows,
\begin{equation}
\hat{\phi}(x)=\int \frac{d^4 p}{(2\pi)^4} ~\! \hat{C}(p) e^{-ipx}.
\label{e0}
\end{equation}
Substituting in eq.(\ref{eqmot}), we get
\begin{equation}
\int d^4p~\! (-\Delta^{\mu\nu}p_\mu p_\nu+m^2)\hat{C}(p) e^{-ipx}=0,
\label{e1}
\end{equation}
from which we find 
\begin{equation}
\hat{C}(p)=\delta(\Delta^{\mu\nu}p_\mu p_\nu-m^2) \hat{a}(p).
\label{e2}
\end{equation}
Now, by using the above result in eq. (\ref{e0}), we get,
\begin{equation}
\hat{\phi}(x)=\int \widetilde{dp}\frac{1}{\sqrt{\Delta^{00}}}
\left( \hat{a}(p)e^{-ipx}+\hat{a}^\dagger(p) e^{ipx}\right),
\label{e4}
\end{equation}
where the time-component $p_0$ is given by
\begin{equation}
p_0=\Delta(p)-\frac{\Delta^{0j}p_j}{\Delta^{00}},\qquad 
\Delta(p)=\sqrt{\left(\frac{\Delta^{0j}p_j}{\Delta^{00}}\right)^2+\frac{m^2-\Delta^{ij}p_ip_j}{\Delta^{00}}},
\label{e5}
\end{equation}
and
\begin{equation}
\widetilde{dp}=\frac{d^3{\bf p}}{(2\pi)^3 2\Delta(p)}.
\label{e6}
\end{equation}
Note that we have also redefined $\hat{a}(p)$ in eq. (\ref{e4}) by a scale factor $\Delta^{00}$, just for convenience. By substituting eq.(\ref{e4}) in eq. (\ref{mconjugate}), we get the corresponding expansion for the conjugate momentum,
\begin{equation}
\hat{\pi}(x)=-i \int \widetilde{dp} \sqrt{\Delta^{00}} \Delta(p)\left( \hat{a}(p)e^{-ipx}-\hat{a}^\dagger(p) e^{ipx}\right).
\label{e7}
\end{equation}
Now, solving for the operators $\hat{a}(p)$ and $\hat{a}^\dagger(p)$, we get
\begin{eqnarray}
\hat{a}(p)&=&i\int\frac{ d^3{\bf x}}{\sqrt{\Delta^{00}}}(\hat{\pi}(x)-i\Delta^{00}\Delta(p)\hat{\phi}(x)) e^{ipx},\nonumber\\
\hat{a}^\dagger(p)&=&-i\int \frac{ d^3{\bf x}}{\sqrt{\Delta^{00}}}(\hat{\pi}(x)+i\Delta^{00}\Delta(p)\hat{\phi}(x)) e^{-ipx},
\label{e8}
\end{eqnarray}
which satisfy the following commutation relations,
\begin{equation}
[\hat{a}(p),\hat{a}^\dagger(q)]=2\Delta (q) (2\pi)^3\delta({\bf p}-{\bf q}),\qquad
[\hat{a}(p),\hat{a}(q)]=[\hat{a}^\dagger(p),\hat{a}^\dagger(q)]=0.
\label{e9}
\end{equation}
From eqs. (\ref{hamiltonian}), 
(\ref{momentum}), (\ref{e4}) and (\ref{e7}), we obtain the  hamiltonian and momentum operators in terms of annihilation and creation operators, namely,
\begin{equation}
\hat{H}=\int \widetilde{dp}~\!\frac{p_0}{2} 
\left( \hat{a}^\dagger(p)\hat{a}(p)+\hat{a}(p)\hat{a}^\dagger(p)\right),
\label{e10}
\end{equation}
and
\begin{equation}
\hat{{\bf P}}=\int \widetilde{dp}~\!\frac{ {\bf p}}{2}
\left(\hat{a}^\dagger(p)\hat{a}(p)+\hat{a}(p)\hat{a}^\dagger(p)\right).
\label{e11}
\end{equation}
Then, we can conclude, from the above expressions and the commutation relations (\ref{e9}), that the spectrum of the theory consists of scalar particles with energy and momentum related by eq. (\ref{e5}). In fact,  we have for the particle energy in terms
of LV  tensor coefficients $\kappa^{\mu\nu}$, the following expression
\begin{equation}
p_0=\sqrt{\left(\frac{\kappa^{0j}p_j}{1+\kappa^{00}}\right)^2+\frac{m^2+{\bf p}^2-\kappa^{ij}p_ip_j}{1+\kappa^{00}}}-
\frac{\kappa^{0j}p_j}{1+\kappa^{00}},
\label{e12}
\end{equation}
from which it is possible notice that the energy is real, positive and bounded from below whenever the second term within the square root is positive. Otherwise, the particle energy spectrum might exhibit unboundedness or complex values. Considering  $\kappa^{00}>-1$,  these issues can be ruled out if the eigenvalues of the matrix $\kappa^{ij}$ are less than the unity, i.e. $|\kappa^{ij}|<<1$, which is naturally expected for a realistic theory.
For other values of $\kappa^{ij}$ it is possible to fix some suitable limits in order to have a well defined particle spectrum. For instance, by
considering the case in which all diagonal terms are equal to $\kappa^{11}$, and respectively all off-diagonal terms are $\kappa^{12}$, we find that the suitable conditions are
$\kappa^{11}-\kappa^{12}<1$, and $\kappa^{11}+2\kappa^{12}<1$. On the other hand, by considering arbitrary diagonal terms, $\kappa^{11}$,
$\kappa^{22}$, $\kappa^{33}$, and all the off-diagonal terms vanishing except for $\kappa^{12}=\kappa^{21}$, we find a well behaved particle spectrum
if $\kappa^{33}<1$, $\kappa^{11}+\kappa^{22}<1$ and $ \kappa^{11}+\kappa^{22}+(\kappa^{12})^2-\kappa^{11}\kappa^{22}<1$.

Thus, in order to avoid those sort of undesirable behaviours for particle spectrum, from now on we will assume that the magnitude of the constant tensor components  $\kappa^{\mu\nu}$ is very small, and proper conditions guaranteeing a well-defined particle spectrum are satisfied.

%%%%%%%%%%%%%%%%%%%%%%%%%%%%%%%%%%%%%%
\section{Thermal behaviour for the free scalar field}
\label{free}
%%%%%%%%%%%%%%%%%%%%%%%%%%%%%%%%%%%%%%
In this section we analyze the effects and contributions of the LV terms on the thermodynamics of the free real scalar field. Let us start from the partition function,
\begin{equation}
Z=\int {\cal D} \phi~\! \exp \left( -{\int}_{\!\!\!0}^{\beta} d\tau \int d^3 {\bf x}~\!{\cal L}_E\right),
\label{e31}
\end{equation}
where $\beta=T^{-1}$ is the inverse of the temperature\footnote{Here we are using the usual convention in which the Boltzmann constant is $k_B=1$, as well as $c=\hbar=1$.}, and ${\cal L}_E$ is the Euclidean density lagrangian of the theory with LV terms, obtained from the prescription $t\to i\tau$, and given by
\begin{equation}
{\cal L}_E=\frac{1}{2}\overline{\Delta}_{\mu\nu}\partial_\mu\phi(x)\partial_\nu\phi(x)+
U(\phi),
\label{e32}
\end{equation}
with $U(\phi)$ contains the mass and possible interaction terms,  and
\begin{equation}
\overline{\Delta}_{00}=1+\kappa^{00},~~\overline{\Delta}_{0j}=-i\kappa^{0j},~~\overline{\Delta}_{ij}=\delta^{ij}-\kappa^{ij}.
\label{e33}
\end{equation}
It is worth pointing out that the integration in Eq. (\ref{e31}) must be computed over all  periodic field configurations, i.e. $\phi({\bf x},\tau+\beta)=\phi({\bf x},\beta)$. 
Now, the thermal Green functions, defined as the thermal expectation values of imaginary time-ordered products of field operators,  are given by
\begin{eqnarray}
\langle \phi(x_1)...\phi(x_n)\rangle&=&\frac{1}{Z}\int {\cal D} \phi \,\, \phi(x_1)...\phi(x_n)\,\exp \left( -{\int}_{\!\!\!0}^{\beta} d\tau \int d^3 {\bf x}~\!{\cal L}_E\right)
 \nonumber\\
&=&\left. \frac{1}{Z(j)}\frac{\delta ^n Z(j)}{\delta j(x_1)...\delta j(x_n)}\right|_{j=0},
\label{gr1}
\end{eqnarray}
where $Z(j)$ denotes the thermal generating functional,
\begin{equation}
Z(j)=\int {\cal D} \phi~\! \exp \left( -{\int}_{\!\!\!0}^{\beta} d\tau \int d^3 {\bf x}~\!({\cal L}_E-j\phi)\right).
\label{gf}
\end{equation}
Firstly, we will focus on the free case,  $U(\phi)=m^2\phi^2/2$. In that case
the associated thermal propagator $ D(x,y)$ for the free real scalar field, 
which corresponds to the two-point Green function (\ref{gr1}), follows the equation,
\begin{equation}
\overline{\square} D(\bf{x},\tau)=\delta(\bf{x})\delta(\tau),
\label{proptef4}
\end{equation}
where $\overline{\square}=\left(-\overline{\Delta}_{\mu\nu}\partial_\mu\partial_\nu+m^2\right) $, and we have taken \mbox{$y=0$} without loss of generality.
Now, considering periodic boundary conditions in the imaginary time, we can write $ D(\bf{x},\tau)$, as follows 
\begin{equation}
D({\bf x},\tau)=\frac{1}{\beta}\sum_{n=-\infty}^\infty\int \frac{d^3{\bf p}}{(2\pi)^3} e^{-i(\omega_n\tau-\bf{p}.\bf{x})}
\widetilde{D}({\bf p},\omega_n),
\label{propterel}
\end{equation}
where $\omega_n=2\pi n/\beta$,  $n=0,\pm 1, \pm 2,...$ are the Matsubara frequencies.
Substituting eq. (\ref{propterel}) in eq. (\ref{proptef4}), we get the thermal propagator in the Fourier space,
\begin{equation}
\widetilde{D}({\bf p},\omega_n)=\frac{1}{\overline{\Delta}_{00}\,\omega_n^2-2\omega_n \overline{\Delta}_{0j}p_j+
\overline{\Delta}_{ij}p_ip_j+m^2}.
\label{thermalprop}
\end{equation}
Notice that this representation of the thermal propagator is complex-valued since $\overline{\Delta} _{0j}$ is imaginary. However, this is not a problem since its representation in the configuration space (\ref{propterel}) is real, as it is expected, since we are dealing with
a real field.

Now, we will compute the partition function  (\ref{e31}), 
from which we will obtain  thermodynamic quantities as the internal energy, pressure, and entropy of the system.
From Eq. (\ref{e31}) we find
\begin{equation}
Z_0=\int {\cal D} \phi ~\!
 \exp \left(-\frac{1}{2}{\int}_{\!\!\!0}^{\beta} d\tau \int  
 d^3 {\bf x} ~\!\phi(x)\overline{\square} \phi(x)   \right),
\label{e34}
\end{equation}
where the zero lower label stands for the free case. As usual, taking the system inside a box of volume $V$, and considering periodic boundary conditions, we can expand the field as follows
\begin{equation}
\phi(\tau, {\bf x})=\sqrt{\frac{\beta}{V}} \sum_{n=-\infty}^\infty \sum_{{\bf p}} \tilde{\phi}_n({\bf p})
e^{-i (\omega_n \tau-{\bf p}.{\bf x})}.
\label{e35}
\end{equation}
Now, by using the orthogonality relations
\begin{equation}
{\int}_{\!\!\!0}^{\beta} d\tau \int d^3{\bf x} ~\!e^{i (\omega_n-\omega_l) \tau-i({\bf p}-{\bf q}).{\bf x}}=\beta V \delta_{n l}~\!\delta_{{\bf p} {\bf q}},
\label{e36}
\end{equation} 
and $\tilde{\phi}_{-n}(-{\bf p})=\tilde{\phi}_n^\ast({\bf p})$, we can write
\begin{equation}
Z_0=N\int \prod_{n}\prod_{ {\bf p}} d\tilde{\phi}_n({\bf p}) \exp\left[ -\frac{\beta^2}{2}\sum_{n=-\infty}^\infty \sum_{\bf p}\widetilde{D}^{-1}({\bf p},\omega_n)   |\tilde{\phi}_n({\bf p})|^2
\right].
\label{e37}
\end{equation}
The integrals above can be computed by writing $\tilde{\phi}_n({\bf p})$ in polar form, and integrating in the corresponding modulus. Doing that, we obtain
\begin{equation}
Z_0=N'\prod_{n{\bf p}}\left[ \beta^2\left(\overline{\Delta}_{00}\omega_n^2-2\overline{\Delta}_{0j}\omega_n p_j+\overline{\Delta}_{ij}p_i p_j+m^2\right)\right]^{-1/2},
\label{e38}
\end{equation}
where contributions from the phase integrals have been incorporated  in $N'$, and therefore we find 
\begin{equation}
\ln Z_0=\ln N'-\frac{1}{2}\sum_{n=-\infty}^\infty \sum_{\bf p} \ln 
 \left(4\pi^2 n^2-4\pi \beta \frac{\overline{\Delta}_{0j}}{\overline{\Delta}_{00}} n p_j+\beta^2\frac{\overline{\Delta}_{ij}}
{\overline{\Delta}_{00}}
p_i p_j+\beta^2\frac{m^2}{\overline{\Delta}_{00}}\right).
\label{e39}
\end{equation}
Now, by summing over $n$, and noting that $N'$ is independent from the LV coefficients, we can proceed by using a similar prescription as in the Lorentz invariant case (see \mbox{appendix \ref{appA})}, and then we have
\begin{equation}
\ln Z_0=-V \int \frac{ d^3 {\bf p}}{(2\pi)^3}\left[\frac{1}{2}\beta p_0+\ln\left(1-e^{-\beta p_0}\right)\right],
\label{e40}
\end{equation}
where $p_0$ given in eq. (\ref{e12}) is the particle energy spectrum. The first term in the above expression is divergent and represent the vacuum energy, which has been
analyzed recently in connection with the Casimir energy in Refs. 
\cite{Cruz2017,Escobar2020,Escobar2,Cruz2020}. 
Since we are interested only in studying thermal properties, we will disregard that term
from now on. We notice also  that as in the LI case,  the expression (\ref{e40}) scales with the
volume $V$, which is a consequence of translational  invariance of the system.   To compute explicitly the expression (\ref{e40}),  we will consider separately the massless and the massive case. 
 %
%%%%%%%%%%%%%%%%%%%%%%%%%%%%%%%%%%%%%%%%%%%%%
\subsection{The free massless case}
%%%%%%%%%%%%%%%%%%%%%%%%%%%%%%%%%%%%%%%%%%%%%
The integral (\ref{e40}) can be computed exactly for the massless case. In this case the particle energy, which we will denote as $\tilde{p}_0$, takes the following form,
\begin{equation}
\tilde{p}_0=\sqrt{p_i A_{ij}p_j}-\frac{\kappa^{0j}p_j}{1+\kappa^{00}},\qquad \quad A_{ij}=\frac{\kappa^{0i}\kappa^{0j}}{(1+\kappa^{00})^2}
+\frac{\delta_{ij}-\kappa^{ij}}{1+\kappa^{00}}.
\label{e41}
\end{equation}
Now, we introduce a matrix $M_{ij}$  that diagonalizes $A_{ij}$, namely $M^{-1}A M= {\rm diag}(a_1,a_2,a_3)$, 
%\label{e42}
%\end{equation}
%
where $a_1, a_2$ and $a_3$, are the eigenvalues of matrix $A_{ij}$. By defining new variables
\begin{eqnarray}
p_i= \frac{1}{\beta\sqrt{a_j}} M_{ij}\bar{p}_j,
\label{novop}
\end{eqnarray}
and disregarding the vacuum energy contribution, we get for (\ref{e40}),
\begin{equation}
\ln Z_0\big|_{m=0}=-V \frac{|\det M|}{|\det A|^{1/2}}\beta^{-3}\int \frac{ d^3 {\bf \bar{p}}}{(2\pi)^3}
\ln\left[1-e^{-\bar{p}+{\bf b}.{\bf \bar{p}} }\right],
\label{e43}
\end{equation}
where $\bar{p}=|{\bf \bar{p}}|$, and the components of ${\bf b}$ are
\begin{equation}
b^{j}=\frac{\kappa^{0i}M_{ij}}{\sqrt{a_j}(1+\kappa^{00})}.
\label{e43a}
\end{equation}
The integral in (\ref{e43}) can be computed straightforwardly in spherical coordinates, obtaining the following final result
\begin{equation}
\ln Z_0\big|_{m=0}=V \frac{\pi^2 }{90(1-b^2)^2} \frac{|\det M|}{|\det A|^{1/2}}\beta^{-3},
\label{e44}
\end{equation}
where $b=|{\bf b}|$. From this result we can determine the internal energy,
\begin{equation}
U=V \frac{\pi^2  }{30(1-b^2)^2} \frac{|\det M|}{|\det A|^{1/2}} T^4,
\label{e45}
\end{equation}
and the pressure, 
\begin{equation}
P=\frac{\pi^2 }{90(1-b^2)^2} \frac{|\det M|}{|\det A|^{1/2}} T^4,
\label{e46}
\end{equation}
from which we obtain the relation,
\begin{equation}
PV=T \ln Z_0\big|_{m=0}\,=\, \frac{U}{3},
\label{e49}
\end{equation}
where the dependence on the LV parameters is hidden. However, as we will see in the next subsection, that behavior is a peculiarity of the massless case.
Now, considering  the mean number of particles
\begin{equation}
{N}=V\int \frac{d^3{\bf p}}{(2\pi)^3}\frac{1}{e^{\tilde{p}_0 \beta}-1},
\label{e47}
\end{equation}
and  using (\ref{e41}), we obtain
\begin{equation}
{N}=V \frac{1}{\pi ^2(1-b^2)^2} \frac{|\det M|}{|\det A|^{1/2}}\zeta(3) T^3.
\label{e48}
\end{equation}
And finally for the entropy, we get
\begin{equation}
S=V \frac{2\pi^2 }{45(1-b^2)^2} \frac{|\det M|}{|\det A|^{1/2}}T^3.
\label{e48a}
\end{equation}
Since all of the above thermodynamics quantities are derived from eq. (\ref{e44}), they are all modified in relation to the Lorentz-invariant case by a global multiplicative factor. Also, from eq. (\ref{e48}) we  see that the mean number of particles is modified by the same factor. From this behaviour, we can conclude in general that for small violating parameters $\kappa$,
the  thermodynamics quantities increase or decrease according to whether these parameters are positive or negative. To illustrate that, we can evaluate explicitly eq. (\ref{e44}) for some specific situations consistent with the conditions discussed at the end of section \ref{spectrum}. In the first scenario, let us consider all the $\kappa$ components as null except for $\kappa^{00}$ and $\kappa^{11}$. Then we obtain
\begin{equation}
   \ln Z_0\big|_{m=0}=V \frac{\pi^2 }{90} \frac{\left(1+\kappa^{00}\right)^{3/2}}{(1-\kappa^{11})^{1/2}}\beta^{-3}.
\end{equation}
In the second scenario, we consider $\kappa^{02}$ as the only non-null component. So, we get,
\begin{equation}
    \ln Z_0\big|_{m=0}=V \frac{\pi^2 }{90}{\left(1+(\kappa^{02})^2 \right)^{3/2} 
}\beta^{-3}.
\end{equation}
%
%Now, considering $\kappa^{ij}=0$, for $i=j$, and $\kappa^{12}$ for all terms in off-diagonal, we have,
%
%\begin{equation}
%\ln Z_0=V \frac{\pi^2 }{90} %\frac{3\left(1+\kappa^{00}\right)^{3/2}}{|1-3 %(\kappa^{12})^{2}-2 (\kappa^{12})^3|^{1/2}}\beta^{-3}.    
%\end{equation}
%
For the last scenario, we choose $\kappa^{12}$ as the only non-null component. Hence,
\begin{equation}
    \ln Z_0\big|_{m=0}=V \frac{\pi^2 }{45} \frac{1}{(1-3 (\kappa^{12})^{2})^{1/2}}\beta^{-3}.
\end{equation}
In general, we notice that the global multiplicative factor, which contain all the corrections, can in principle increase or decrease the standard result, i.e. the one obtained in the Lorentz-invariant case, by a linear or quadratic contribution of the $\kappa$ components, depending on whether they are diagonal or off-diagonal terms. Of course, by considering that all the $\kappa$ components are non-null, we will obtain a more complicated form which will contain higher powers of them. However, those additional terms will not represent any significant contributions. In fact, if we expand the multiplicative factor in powers of these components, it will be enough to keep up to linear order for practical purposes.

%%%%%%%%%%%%%%%%%%%%%%%%%%%%%%%%%%%
\subsection{The free massive case}
%%%%%%%%%%%%%%%%%%%%%%%%%%%%%%%%%%%
Now, we will evaluate the partition function for the free massive case. Since we are not able to perform exactly the integral (\ref{e40}), we are going to consider the high temperature regime instead, {\it i.e.}
$m/T<<1$ ($m\b<<1$), and then we will expand the thermodynamics quantities in terms of this parameter. First of all, we expand
$p_0$ (\ref{e12}) as follows
\begin{equation}
p_0 \simeq  \tilde{p}_0+\delta p_0,\qquad \mbox{with} \qquad \delta p_0=\frac{m^2}{2(1+\kappa^{00})\sqrt{p_i A_{ij}p_j}},
\label{e52}
\end{equation}
where $\tilde{p}_0$ and $A_{ij}$ are given in eq.(\ref{e41}). Substituting the above approximation in the integral (\ref{e40}), and expanding 
up to first order in $\delta p_0$, we have
\begin{equation}
\ln Z_0=\ln Z_0\big|_{m=0}-V\beta\int \frac{d^3{\bf p}}{(2\pi)^3}\frac{\delta p_0}{e^{\beta \tilde{p}_0}-1}+....
\label{e53a} 
\end{equation}
Now, by using the redefinition (\ref{novop}), we obtain
\begin{equation}
\ln Z_0=\ln Z_0\big|_{m=0}-V\beta^{-1}m^2\frac{|\det M|}{2|\det A|^{1/2} (1+\kappa^{00})}\int \frac{d^3{\bf \bar{p}}}{(2\pi)^3}
\frac{1}{\bar{p}(e^{{\bar p}-{\bf b}.{\bf \bar{p}}}-1)}+....
\label{e53}
\end{equation}
Now, performing the final integration, we finally get,
\begin{eqnarray}
\ln Z_0
\!&=&\!V \frac{\pi^2 }{90(1-b^2)^2} \frac{|\det M|}{|\det A|^{1/2}}\beta^{-3} \left( 1-\frac{15(1-b^2)}{4\pi^2(1+\kappa^{00})}\beta^2m^2+
{\cal O}((\beta m)^4)
\right).
\label{e54}
\end{eqnarray}
%%%%%%%%%%%
%%%%%%%%%%%
As it was done for the massless case,  from the above result we can determine the internal energy and pressure,  namely
\begin{eqnarray}
U\!&=&\! V \frac{\pi^2 }{30(1-b^2)^2} \frac{|\det M|}{|\det A|^{1/2}}T^4\left( 1-\frac{5(1-b^2)}{4\pi^2(1+\kappa^{00})}m^2T^{-2}+
{\cal O}((m/T)^4)
\right),
\label{e55}\\
P\!&=&\! \frac{\pi^2 }{90(1-b^2)^2} \frac{|\det M|}{|\det A|^{1/2}}T^4 \left( 1-\frac{15(1-b^2)}{4\pi^2(1+\kappa^{00})}m^2 T^{-2}+
{\cal O}((m/T)^4)\right),
\label{e56}
\end{eqnarray}
which satisfy the relation
\begin{equation}
PV=T \ln Z_0 \, =\, \frac{U}{3} \left( 1-\frac{10(1-b^2)}{4\pi^2(1+\kappa^{00})}m^2 T^{-2}+
{\cal O}((m/T)^4)\right).
\label{e56a}
\end{equation}
%
%Clearly, the above relation is a generalization of  (\ref{e49}), but unlike the massless case, the Lorentz-violating parameters appear
%explicitly.  A
Also,  we find that the entropy is given by
\begin{equation}
S=V \frac{\pi^2 }{90(1-b^2)^2} \frac{|\det M|}{|\det A|^{1/2}}T^3\left( 4-\frac{15(1-b^2)}{2\pi^2(1+\kappa^{00})}m^2T^{-2}+
{\cal O}((m/T)^4)
\right)    ,
\label{e56b}
\end{equation}
and  the mean number of particles reads
\begin{equation}
N=  \frac{ V |\det M| T^3 }{\pi^2 |\det A|^{1/2}}\left(  \zeta(3) +\frac{m^2 T^{-2}}{4(1+\kappa^{00})}
\ln\sinh\left(   \frac{m}{2\sqrt{(1+\kappa^{00})}T}\right)
+{\cal O}(( m/T)^4)
\right),
\label{e57}
\end{equation}
where it has been considered  only the case $b=0$, for simplicity (see appendix \ref{appB} for more details). 

Now, as it was done in the massless case, we will analyze the effect of LV terms on the expression (\ref {e54}). Let us first consider that only $\kappa^{00} \neq 0$, and the rest of the components are null. Up to
second order in $\beta m$, we find,
\begin{equation}
    \ln Z_0=V \frac{\pi^2 }{90}\left(1+\kappa^{00}\right)^{3/2}\beta^{-3} \left( 1-\frac{15}{4\pi^2(1+\kappa^{00})}\beta^2m^2 \right).
\end{equation}
If we consider that only the $\kappa^{11}$ component is different from zero, we obtain
\begin{equation}
    \ln Z_0=V \frac{\pi^2 }{90}\left(1+\kappa^{11}\right)^{-1/2}\beta^{-3} \left( 1-\frac{15}{4\pi^2}\beta^2m^2 \right).
\end{equation}
We note that both  results suffer an explicit global effect from the LV parameters, but only the term $\kappa^{00}$ appears explicitly in the term proportional to mass. For $0<\kappa^{00}<1$, the global term of pressure and energy increase, but the term proportional to the mass decreases these quantities. For $-1<\kappa^{00}<0$, the global and the mass terms decrease the pressure and energy. Comparing these two cases, we can see that for the positive $\kappa^{00}$, the mass term decrease these thermodynamics quantities more than the negative $\kappa^{00}$ case.
In the case of $\kappa^{11} \neq 0$, the Lorentz-violation term appears only as a global term. For negative $\kappa^{11}$, the pressure and energy decreases more than the positive $\kappa^{11}$ case. On the other hand, when considering the cross-terms $\kappa^{0i}$ and $\kappa^{ij}$, we conclude that they appear, at least, in quadratic order in the equation (\ref{e54}). Regarding the state equations, we see that, unlike the massless case, there is an explicit effect of the LV terms, even when written in terms of $N$.

%%%%%%%%%%%%%%%%%%%%%%%%%%%%%%%%%%%%%%%%%%%%%
\section{The thermal behaviour for the interacting scalar field }
\label{interacting}
%%%%%%%%%%%%%%%%%%%%%%%%%%%%%%%%%%%%%%%%%%%%%%
In this section we consider the study of the thermal properties of the real scalar interacting field, with euclidean lagrangian given by (\ref{e32}), where
\begin{equation}
U(\phi)=\frac{m^2}{2}\phi^2+\frac{\lambda}{4!}\phi^4.
\label{denspot}
\end{equation}
To compute the partition function perturbatively, we decompose the euclidean action in the free and interacting part,
\begin{equation}
S_E=S_0+S_I,
\label{e60}
\end{equation} 
where $S_0$ is the free euclidean action and $S_I$ is given by the integral of the last term in (\ref{denspot}). 
In this way, expanding the interacting part we have from (\ref{e31}),
\begin{equation}
\ln Z=\ln Z_0+\ln Z_I,
\label{e61}
\end{equation}
where $Z_0$ is the free partition function considered and evaluated  in the last section,  and $\ln Z_I$ is given by,
\begin{eqnarray}
\ln Z_I&=&\ln\left( 1+\frac{1}{Z_0}\sum_{n=1}^\infty\frac{(-1)^n}{n!} \int {\cal D} \phi ~\!e^{-S_0} (S_I)^n\right)
\nonumber\\
&=& \ln \left(
1+\sum_{n=1}^\infty \frac{(-\lambda)^n}{4^n n!} \int dx_1...\int dx_n \langle \phi^4(\tau_1,{\bf  x}_1)...\phi^4(\tau_n,{\bf x}_n)\rangle
\right).
\label{e62}
\end{eqnarray}
In above expression, we have used the shorthand notation $dx$ to denote $d\tau d^3{\bf x}$ and
\begin{equation}
\langle ...\rangle=\frac{1}{Z_0}\int {\cal D} \phi~\! ... e^{-S_0},
\label{e63}
\end{equation}
that we can compute using the Wick theorem with the help of the free propagator (\ref{propterel})-(\ref{thermalprop}). In this way it
is possible to show that (\ref{e62}) is given schematically  by the connected  vacuum graphs,  but replacing the zero temperature propagator by  the thermal propagator, and also the vertex $-i\lambda$ by  $-\lambda$. So, at first order in the coupling constant, eq. (\ref{e62}) reads
\begin{eqnarray}
\ln Z_I
&=&-\frac{\lambda}{4!}\int_0^\beta d\tau \int d^3{\bf x} \, \langle \phi^4(x)\rangle
\nonumber\\
&=&-\frac{\lambda}{8} \int_0^\beta d\tau \int d^3{\bf x}\,  [D({\bf 0},0)]^2.
\label{e64}
\end{eqnarray}
Using (\ref{propterel}) and (\ref{thermalprop}),  the above expression can be written as
\begin{equation}
    \ln Z_I=-V\frac{\lambda}{8\beta}\left(\sum_{n=-\infty}^\infty
    \int \frac{d^3{\bf p}}{(2\pi)^3} 
    \frac{1}{\overline{\Delta}_{00}\,\omega_n^2-2\omega_n \overline{\Delta}_{0j}p_j+
\overline{\Delta}_{ij}p_ip_j+m^2}
    \right)^2.
    \label{e65}
\end{equation}
Now, by replacing the Matsubara frequencies in (\ref{e65}), and summing  in $n$, we get  
\begin{eqnarray}
  \ln Z_I&=&- V\beta\frac{\lambda  }{8}\left[
  \int \frac{d^3{\bf p}}{(2\pi)^3}
  \big( {\cal F}_{+} (p) +{\cal F}_{-} (p)\big)
  \right]^2,
  \label{equ67}
\end{eqnarray}
where
\br 
 {\cal F}_{\pm} (p)= \frac{1}  {4{\cal R}}\coth\bigg(\frac{\beta}{2\overline{\Delta}_{00}}\Big({\cal R} \pm 
  i\overline{\Delta}_{0j}p_j\Big)\bigg),
\er
with
\br 
{\cal R }= \sqrt{(i\overline{\Delta}_{0j}p_j)^2  +\overline{\Delta}_{00}(\overline{\Delta}_{ij}p_ip_j+m^2)}.
\er
By performing the transformation ${\bf p}\to -{\bf  p}$,  we can show that $ {\cal F}_{+}$ and $ {\cal F}_{-}$ within the integral  (\ref{equ67}) are equal, and then using (\ref{e33}) we get
\begin{eqnarray}
     \ln Z_I&=&-V\beta  \frac{\lambda}{32}\left[
  \int \frac{d^3{\bf p}}{(2\pi)^3}
  \frac{\coth(\beta p_0/2)}
  {(1+\kappa^{00})p_0+\kappa^{0j}p_j}
  \right]^2\nonumber\\
 % \label{e66}
&=&-V\beta  \frac{\lambda}{32}\left[
  \int \frac{d^3{\bf p}}{(2\pi)^3}
  \frac{1}
  {(1+\kappa^{00})p_0+\kappa^{0j}p_j}
  + \frac{2}
  {[(1+\kappa^{00})p_0+\kappa^{0j}p_j](e^{\beta p_0}-1)}
  \right]^2,\qquad \mbox{}
  \label{e68} 
\end{eqnarray}
where $p_0$ is the free particle energy, given by (\ref{e12}). The first term in (\ref{e68}), is ultraviolet divergent but linear in the  inverse temperature $\beta$. Such term gives  the first-order vacuum energy correction and has been treated recently in connection to the 
radiative corrections to the Casimir energy \cite{MMZ}.   Also,  since we are interested  
only in  studying the thermal properties of the system, from now on we disregard such contribution  as we did in the free case. In this way, we get for
the finite temperature dependent part,
\begin{equation}
  \ln Z_I=-V\beta  \frac{\lambda}{8}\left[
  \int \frac{d^3{\bf p}}{(2\pi)^3}
  \frac{1}
  {[(1+\kappa^{00})p_0+\kappa^{0j}p_j](e^{\beta p_0}-1)}
  \right]^2.
  \label{e69}    
\end{equation}
%
%%%%%%%%%%%%%%%%%%%%%%%%%%%%%%%%%
\subsection{The massless case}
%%%%%%%%%%%%%%%%%%%%%%%%%%%%%%%%%%
For the massless case, the above integral  can be done exactly. In this case, by using
(\ref{e41}) in (\ref{e69}), and performing the change of variable (\ref{novop}), we have
\begin{equation}
\ln Z_I{\big |}_{m=0}=-V\beta \frac{\lambda}{8}
\left[\frac{\beta^{-2}|\det M|}{|\det A|^{1/2}(1+\kappa^{00})}
\int \frac{d^3{\bf \bar{p}}}{(2\pi)^3}\frac{1}
{\bar{p}(e^{\bar{p}-{\bf b}.{\bf \bar{p}}}-1)}
\right]^2.
\label{e70}
\end{equation}
Now, using polar coordinates, we obtain
\begin{equation}
\ln Z_I{\big |}_{m=0}=-V\beta^{-3} \frac{\lambda}{2^7 \cdot 9}
\frac{|\det M|^2}{|\det A|(1+\kappa^{00})^2(1-b^2)^2}.
\label{e71}   
\end{equation}
Replacing eqs. (\ref{e44}) and (\ref{e71}) in eq. (\ref{e61}), we have for the massless interacting partition function
\begin{equation}
\ln Z {\big |}_{m=0}=V  \frac{|\det M|}{(1-b^2)^2|\det A|^{1/2}}\beta^{-3}
\left(\frac{\pi^2 }{90} - \frac{\lambda}{2^7\cdot 9}
\frac{|\det M|}{|\det A|^{1/2}(1+\kappa^{00})^2}+...
\right),
\label{e72}
\end{equation}
from which the internal energy and pressure can be obtained,
\begin{equation}
U= V  \frac{|\det M|}{(1-b^2)^2|\det A|^{1/2}}T^{4}
\left(\frac{\pi^2 }{30} - \frac{\lambda}{2^7\cdot 3}
\frac{|\det M|}{|\det A|^{1/2}(1+\kappa^{00})^2}+...
\right)  , 
\label{e73}
\end{equation}
and
\begin{equation}
P=  \frac{|\det M|}{(1-b^2)^2|\det A|^{1/2}}T^{4}
\left(\frac{\pi^2 }{90} - \frac{\lambda}{2^7\cdot 9}
\frac{|\det M|}{|\det A|^{1/2}(1+\kappa^{00})^2}+...
\right).   
\label{e74}
\end{equation}
Also, for the entropy, we get
\begin{equation}
S=V  \frac{|\det M|}{(1-b^2)^2|\det A|^{1/2}}T^{3}
\left(\frac{2\pi^2}{45}  - \frac{\lambda}{2^5\cdot 9}
\frac{|\det M|}{|\det A|^{1/2}(1+\kappa^{00})^2}+...
\right). 
\label{e75}
\end{equation}
From (\ref{e73}) and (\ref{e74}) we see that at first order in the coupling constant $PV=U/3$, as
in the free massless case. 

We will now consider some specific cases. When only $\kappa^{00} \neq 0$, it can be noticed that the behavior of the above thermodynamics quantities is similar to the massive case without interaction, where $\lambda$ plays the role of the mass. However, each one of these effects has different magnitude. On the other hand, by considering only $\kappa^{11} \neq 0$, unlike the massive case without interaction, the thermodynamics quantities decrease by the interaction term $\lambda$. However, the similarity with the massive case without interaction comes from the global term that always decreases the pressure and the energy.
%%%%%%%%%%%%%%%%%%%%%%%%%%%%%%%%%%%%%%%%%%%
\subsection{The massive case}
%%%%%%%%%%%%%%%%%%%%%%%%%%%%%%%%%%%%%%%%%%%%
In this situation the integrand in (\ref{e69}) can not be performed exactly. As in the free massive case we expand in the parameter $\beta m$, and then our result will be valid for small values of $\beta m$ (see the full details of the calculation in Appendix \ref{appC}). According
to the free case, we only consider the first-order correction, and then we have for the
interacting massive partition function,
\begin{eqnarray}
\ln Z=&&\!\!\!\!\!\!\!\!V \frac{\pi^2 }{90(1-b^2)^2} \frac{|\det M|}{|\det A|^{1/2}}\beta^{-3} \left( 1-\frac{15(1-b^2)}{4\pi^2(1+\kappa^{00})}\beta^2m^2+
{\cal O}((\beta m)^4).
\right)\nonumber\\
&&\!\!\!\!\!\!\!\!
-V \frac{\lambda}{2^7\cdot 9}
\frac{|\det M|^2}{(1-b^2)^2|\det A|(1+\kappa^{00})^2}\beta^{-3}
\left(1-3\frac{(1-b^2)\tilde{b} }{\pi(1+\kappa^{00})^{1/2}}\beta m
+...\right),
\label{e76}
\end{eqnarray}
where $\tilde{b}$ is given by
\begin{equation}
\tilde{b}=(1+b)^{-1/2}+(1-b)^{-1/2}.
\label{e77}
\end{equation}
In the second line in (\ref{e76}) we have disregarded quadratic terms in
$\beta m$,  since being the constant coupling $\lambda$ sufficiently small, we 
expect that crossing terms of the type $\lambda (\beta m)^2$ are also very small 
compared with the free contribution. Now, from (\ref{e76}) we get for the internal energy, the  pressure and entropy, the following expressions
\begin{eqnarray}
U=&&\!\!\!\!\!\!\!\! V \frac{\pi^2 }{30(1-b^2)^2} \frac{|\det M|}{|\det A|^{1/2}} T^4 \left( 1-\frac{5(1-b^2)}{4\pi^2(1+\kappa^{00})}m^2T^{-2}+...
\right)\nonumber\\
&&\!\!\!\!\!\!\!\!
-V \frac{\lambda}{2^7\cdot 3}
\frac{|\det M|^2}{(1-b^2)^2|\det A|(1+\kappa^{00})^2}T^4
\left(1-2\frac{(1-b^2)\tilde{b} }{\pi(1+\kappa^{00})^{1/2}} m T^{-1}
+...\right),
\label{e78}   
\end{eqnarray}
\begin{eqnarray}
P=&&\!\!\!\!\!\!\!\!\frac{\pi^2 }{90(1-b^2)^2} \frac{|\det M|}{|\det A|^{1/2}}T^4
\left( 1-\frac{15(1-b^2)}{4\pi^2(1+\kappa^{00})}m^2T^{-2}+
...
\right)\nonumber\\
&&\!\!\!\!\!\!\!\!
- \frac{\lambda}{2^7\cdot 9}
\frac{|\det M|^2}{(1-b^2)^2|\det A|(1+\kappa^{00})^2}T^4
\left(1-3\frac{(1-b^2)\tilde{b}}{\pi(1+\kappa^{00})^{1/2}} m T^{-1}
+...\right),
\label{e79}    
\end{eqnarray}
and
\begin{eqnarray}
S=&&\!\!\!\!\!\!\!\!V \frac{2\pi^2 }{45(1-b^2)^2} \frac{|\det M|}{|\det A|^{1/2}} T^3 \left( 1-\frac{15(1-b^2)}{8\pi^2(1+\kappa^{00})}m^2T^{-2}+...
\right)\nonumber\\
&&\!\!\!\!\!\!\!\!
-V \frac{\lambda}{2^5\cdot 9}
\frac{|\det M|^2}{(1-b^2)^2|\det A|(1+\kappa^{00})^2}T^3
\left(1-\frac{9}{4}\frac{(1-b^2)\tilde{b} }{\pi(1+\kappa^{00})^{1/2}} m T^{-1}
+...\right).
\label{e80}    
\end{eqnarray}
We notice from these results that when $\kappa^{00}$ is the only non-null coefficient, and \mbox{$-1 < \kappa^{00} < 0$}, the global factor decrease the thermodynamics quantities. On the contrary, when \mbox{$0 < \kappa^{00} < 1$}, the global factor will  increases  these quantities. In its turn, the coefficient $\beta^2 m^2$ will gives also a decreasing contribuition, whereas the term proportional to $\lambda \beta m$ will give an increasing contribution. However, the magnitude of such contributions will depend on the sign of $\kappa^{00}$.

On the other hand, by considering that $\kappa^{11}$ is the only non-vanishing LV coefficient, we see that in the regime
$-1 < \kappa^{11} < 0$ the global factor  will decrease the thermodynamics quantities, whereas for $0 < \kappa^{11} < 1$, it will give and increasing contribution. In addition,  the coefficient of  proportional to $\l$ will decrease the thermal quantities, whereas an increasing contribution  will be obtained from the term proportional to $\l\b m$. Both of these contributions will depend on the sign of $\kappa^{11}$.

We can also note that in the case of $\kappa^{00} \neq 0$, there is no explicit LV contribution coming from the term proportional to $\l$. While in the case of  $\kappa^{11} \neq 0$, this same feature occurs for the term proportional to $\beta^2 m^2$.

%%%%%%%%%%%%%%%%%%%%%
%%%%%%%%%%%%%%%%%%%%%%%%%%%
%%%%%%%%%%%%%%%%%%%%%%%%%%%%%%%%%%%%%%

\section{Concluding remarks}
\label{conclusao}

In this paper, we have investigated Lorentz-violation (LV) effects on the thermodynamics properties of a real scalar field theory due to the presence of a constant background tensor field.

In order to find the effects of the Lorentz symmetry breaking terms on our model, we firstly looked for the particle spectrum of the theory, given explicitly in eq. (\ref{e12}), from which we conclude that there should be specific bounds over the values of the $\kappa^{\mu \nu}$ coefficients in order to have a real positive energy that is also bounded from below. Then, we assumed that the magnitude of the constant tensor components are very small, i.e. $|\kappa|<<1$, in all the analysis performed, which is quite reasonable.

We  have also studied the corresponding effects over relevant thermodynamics quantities, like the internal energy, pressure, and the entropy. First of all, we considered the free massless scalar field, with some specific  non-vanishing  coefficients for simplicity. We noticed that the effects over the thermodynamics quantities are all enclosed in a global multiplicative factor, which increases the thermodynamics quantities when the LV coefficients are positive, and decrease them when they take negative values. In particular, we found  that  off-diagonal non-null coefficients yield only  quadratic contributions in the LV coefficients, and thus they have been neglected for our analysis.

When the massive scalar field is considered, we face in general a difficulty in order to evaluate analytically the corresponding partition function, and then only the high temperature regime $m/T<<1$ was considered. In this approximation, we have computed the corresponding thermodynamics quantities with very good accuracy. In this case, we get not only the global factor contribution, but also  a contribution from the coefficient of $m^2$. For instance, when  $0 < \kappa^{00} < 1$, the global factor increase the thermal quantities, whereas the square mass coefficient produces a decrease. On the other hand, when $-1 < \kappa^{00} <0$, both contributions cause the pressure, internal energy and entropy to decrease. In addition, by considering only $\kappa^{11} \neq 0$, we only obtain a purely global contribution, such that when $\kappa^{11} < 0$, the pressure, internal energy, and entropy decrease more rapidly than in the case of positive  $\kappa^{11}$.
 
Subsequently, we analysed the case of a LV massless scalar field in the presence of a $\phi^4$ interaction potential. Interestingly, we noticed that this case is somehow similar to the free massive case, where the coupling constant $\l$ plays the role of the mass $m$, but with LV contributions of different magnitude. This similarity can be understood from the effect of the global factor on the thermal quantities. However, unlike the free massive case, the thermal quantities will decrease because of the interaction factor when we consider $\kappa^{11} \neq 0$.

Finally, we analysed the massive interacting scalar field theory in the LV tensor background. In this case, we find several possibilities of decreasing and increasing contributions coming from the LV coefficients. We noticed that when only $\kappa^{00} \neq 0$, there is no such Lorentz breaking coming from the term proportional to the coupling constant $\l$. While in the case when only  $\kappa^{11} \neq 0$, this same features occur for the term proportional to $\beta^2 m^2$.

We believe that the extension our results to more general LV models is an interesting issue that can be addressed in next works. In particular, it would be interesting to analyse low temperature behaviours, and also extensions to LV models with boundaries. Potential applications, such as  LV effects on cosmological problems \cite{Kanno,Avelino,Arianto,Blas,Armendariz}, or in perturbative quantum field theory \cite{Visser}, would also be of great interest. Authors aim to pursue these further studies in future investigations.

%%%%%%%%%%%%%%%%%%%%%%%%%%%%%%%%%%%%%%%%%%%%%%%%

\mbox{}\\[1cm]
%%%%%%%%%%%%%%%%%%%%%%%%%%%%%
\noindent
{\bf Acknowledgements}
\vskip 0.4cm
\noindent
{A.R.A and G.F.H. thank to the Brazilian agency CAPES for partial financial support.
E.S.S. thanks CAPES for full financial support, and R.G.R. thanks to the Brazilian research agency CNPq for full financial support. }

%%%%%%%%%%%%%%%%%%%%%%%%%%%%%%5

\appendix
\section{Explicit calculation of the partition function}
\label{appA}
Let us start from the expression (\ref{e38}) for the partition function for the free scalar field at finite temperature, namely
\begin{equation}
\label{Z0}
Z_0=N'\prod_{np}\left[\beta^2\Big(\overline{\Delta}_{00}\omega_n^2+2\overline{\Delta}_{0j}\omega_np_j+\overline{\Delta}_{ij}p_ip_j+m^2\Big)\right]^{-1/2},
\end{equation}
from which we straightforwardly obtain,
\begin{eqnarray}
\nonumber
\ln{Z_0} &=& \ln{N'}-\frac{1}{2}\sum_{n=-\infty}^{\infty}\sum_{p}\ln\Big(\beta^2\omega_n^2+2\frac{\overline{\Delta}_{0j}}{\overline{\Delta}_{00}}\beta^2 \omega_n p_j+\frac{\beta^2 \omega^2}{\overline{\Delta}_{00}}\Big),\label{lnZ0}
\end{eqnarray}
where we have disregarded the term that does not depend on the thermodynamics variables, and
\begin{equation}
w^2=\overline{\Delta}_{ij}p_ip_j+m^2,  \qquad \omega_n = \frac{2 \pi n}{\beta}.\label{omega}
\end{equation}
By using (\ref{omega}), and introducing the following notation
\begin{equation}
A=\frac{\kappa_{0j}}{\overline{\Delta}_{00}}\beta p_j, \qquad \theta = \frac{\beta^2 \omega^2}{\overline{\Delta}_{00}},
\end{equation}
the expression can be simplified to the following form
\begin{equation}
\label{lnZ03}
\ln{Z_0} = \ln{N'}-\frac{1}{2}\sum_{n=-\infty}^{\infty}\sum_{p}\ln\Big(4 \pi^2 n^2-4 \pi i A n+\theta\Big).
\end{equation}
Now, let us derive and integrate the right-hand side of (\ref{lnZ03}) with respect to $\theta$. Doing that, we note that the summation can be expressed as follows,
\begin{equation}
\label{int&deremtheta}
\sum_{n=-\infty}^{\infty}\sum_{p}\int{\frac{d\theta}{\left[4 \pi^2 n^2 - 4 \pi i A n + \theta\right]}}=\sum_{p}\int{d\theta\sum_{n=-\infty}^{\infty} \frac{1}{\left[4 \pi^2 n^2 - 4 \pi i A n + \theta\right]}}.
\end{equation}
The summation in  (\ref{int&deremtheta}) can be computed exactly, and the result is given by%
{\small
\begin{eqnarray}
\sum_{n=-\infty}^{\infty}\frac{1}{\left[4 \pi^2 n^2 - 4 \pi i A n + \theta\right]} =  \frac{1}{4\sqrt{A^2+\theta}}\bigg[\coth{\bigg(\frac{A+\sqrt{A^2+\theta}}{2}\bigg)}-\coth{\bigg(\frac{A-\sqrt{A^2+\theta}}{2}\bigg)}\bigg]. \label{soma}
\end{eqnarray}
}
Now, by using eq. (\ref{soma}), we can solve the integral in eq. (\ref{int&deremtheta}),
\begin{eqnarray}
\int{\!\!d\theta\!\sum_{n=-\infty}^{\infty} \frac{1}{\left[4 \pi^2 n^2 - 4 \pi i A n + \theta\right]}}
&=& \ln\bigg[\sinh\bigg(\frac{A-\sqrt{A^2+\theta}}{2}\bigg)\bigg]\nonumber\\&&+\ln\bigg[\sinh\bigg(\frac{A+\sqrt{A^2+\theta}}{2}\bigg)\bigg]+C_1,\label{intsoma}
\end{eqnarray}
where $C_1=C_1(\kappa,\beta)$ is an integration constant that can be used to cancel out the term $\ln N'$ in eq. (\ref{lnZ0}). Then, we find
\begin{eqnarray}
\nonumber
\ln{Z_0} \!&\!=\!&\! -\frac{1}{2} \sum_p \bigg\{\beta\sqrt{\bigg(\frac{\kappa_{0j}}{\overline{\Delta}_{00}}p_j\bigg)^2+\frac{\omega^2}{\overline{\Delta}_{00}}}+\ln\left[1-e^{-\beta\left(\sqrt{\left(\frac{\kappa_{0j}}{\overline{\Delta}_{00}}p_j\right)^2+\frac{\omega^2}{\overline{\Delta}_{00}}}-\frac{\kappa_{0j}}{\overline{\Delta}_{00}}p_j\right)}\right]\\
\label{lnZ05}
\!&\!\!&\!\qquad\qquad+\ln\left[1-e^{-\beta\left(\sqrt{\left(\frac{\kappa_{0j}}{\overline{\Delta}_{00}}p_j\right)^2+\frac{\omega^2}{\overline{\Delta}_{00}}}+\frac{\kappa_{0j}}{\overline{\Delta}_{00}}p_j\right)}\right]\bigg\}.
\end{eqnarray}
Now, we can relabel the variable $p \rightarrow -p$ in the last term of (\ref{lnZ05}), and then simplify the above expression as follows
\begin{eqnarray}
\ln{Z_0} \!&\!=\!&\!\!\! \sum_p \Bigg\{\!-\frac{\beta}{2}\sqrt{\left(\frac{\kappa_{0j}}{\overline{\Delta}_{00}}p_j\right)^2+\frac{\omega^2}{\overline{\Delta}_{00}}}
\label{lnZ06}
-\ln\left[1-e^{-\beta\left(\sqrt{\left(\frac{\kappa_{0j}}{\overline{\Delta}_{00}}p_j\right)^2+\frac{\omega^2}{\overline{\Delta}_{00}}}-\frac{\kappa_{0j}}{\overline{\Delta}_{00}}p_j\right)}\right]\!\!\Bigg\}.
\end{eqnarray}
For the continuous case we will find 
\begin{eqnarray}
\ln{Z_0} \!&\!=\!&\!\!\! -V\!\!\int\!\!\frac{d^3p}{(2\pi)^3} \Bigg\{\!\frac{\beta}{2}\sqrt{\left(\frac{\kappa_{0j}}{\overline{\Delta}_{00}}p_j\right)^2+\frac{\omega^2}{\overline{\Delta}_{00}}}
\label{lnZ0continuo}
+\ln\left[1-e^{-\beta\left(\sqrt{\left(\frac{\kappa_{0j}}{\overline{\Delta}_{00}}p_j\right)^2+\frac{\omega^2}{\overline{\Delta}_{00}}}-\frac{\kappa_{0j}}{\overline{\Delta}_{00}}p_j\right)}\right]\!\!\Bigg\}. \qquad
\end{eqnarray}
%%%%%%%%%%%%%%%%%%%%%%%%%%%%%%%%%%%%%%%%%%%%%%%

\section{Mean number of particle for the massive case}
\label{appB}

Let us consider the mean number of particles for the massive case, given by
\begin{equation}
    N=V \int \frac{d^3 p}{(2\pi)^3}\frac{1}{e^{\beta p_0}-1},
\end{equation}
where $p_0$ is given by equation (\ref{e12}). By using eqs. (\ref{e41}), (\ref{novop}), and considering $\kappa^{0i}=0$ for simplicity, we get
\begin{equation}
    N=\frac{ V |\det M| \beta^{-3} }{ 2 \pi^2 |\det A|^{1/2}} \int d{\bar p} \frac{{\bar p}^2}{e^{\sqrt{{\bar p}^2+a^2}}-1},
    \label{eB3}
\end{equation}
where $\bar{p}$ is defined as in (\ref{novop}) and 
\begin{equation}
a^2=\frac{\beta^2 m^2}{(1+\kappa^{00})}.
\label{eb3a}
\end{equation}
To properly calculate the integral in (\ref{eB3}), valid for small $\beta m$, 
we can expand the integrand in orders of $a$. However, it is not possible to use Taylor's expansion since the integrand is not an analytic function on $a=0$, as one can see taking the first derivative
of (\ref{eB3}).   However, we can sort out this problem by using the identity \cite{Jackiw2}, 
\begin{equation}
    \frac{1}{e^{y}-1}=-\frac{1}{2}+\sum^{\infty}_{n=-\infty}\frac{y}{{y}^2 + 4\pi^2 n^2},\label{adb1}
\end{equation}
in the integral of eq. (\ref{eB3}),
we have
\begin{eqnarray}
 I(a)&=& \int d{\bar p} \frac{{\bar p}^2}{e^{\sqrt{{\bar p}^2+a^2}}-1}\nonumber\\
&=&   
  -\int d{\bar p} \,\,\frac{{\bar p}^2}{2} + \sum^{\infty}_{n=-\infty} \int d{\bar p} \frac{{\bar p}^2 \sqrt{{\bar p}^2 + a^2}}{{\bar p}^2 + a^2 +4\pi^2 n^2} +\delta I,\label{eB16}
\end{eqnarray}
where $\delta I$ in (\ref{eB16}) is a necessary counterterm to ensure convergence of the integral $I(a)$, since the identity (\ref{adb1}) generates a fictitious divergence. Separating the massless terms, we can write
\begin{equation}
    I(a)=I(0) +\delta I + \sum^{\infty}_{n=-\infty}
    \left[ \int d{\bar p} \frac{{\bar p}^2 \sqrt{{\bar p}^2 + a^2}}{{\bar p}^2 +a^2 +4\pi^2 n^2} - \int d{\bar p} \frac{{\bar p}^3}{{\bar p}^2+4\pi^2 n^2} \right].
    \label{eB5}
\end{equation}
Hence, we are concerned with calculating the last two integrals of equation (\ref{eB5}), which we will denote as
\begin{equation}
    J=\int d{\bar p} \frac{{\bar p}^2 \sqrt{{\bar p}^2 + a^2}}{{\bar p}^2 +a^2 +4\pi^2 n^2} \qquad 
    \mbox{and} \qquad K=\widetilde{K} +\int d{\bar p} \frac{{\bar p}^3}{{\bar p}^2+4\pi^2 n^2}.
\end{equation}
where $\widetilde{K}$ is a divergent mass-independent constant, whose derivative arises from the term $n\!=\!0$. Let us begin with $J$. First, we approximate in the following way,
\begin{equation}
    J \approx \int^{\infty}_{0} d{\bar p} \frac{{\bar p}^3}{{\bar p}^2+a^2 +4\pi^2 n^2} + \frac{a^2}{2}\int^{\infty}_{0} d{\bar p} \frac{{\bar p}}{{\bar p}^2+a^2 +4\pi^2 n^2},
\end{equation}
and use dimensional regularization by inserting a parameter $\delta$, as follows
\begin{eqnarray}
    J &\approx& \int^{\infty}_{0} d{\bar p} \frac{{\bar p}^{3-2\delta}}{{\bar p}^2+a^2 +4\pi^2 n^2} + 
    \frac{a^2}{2}\int^{\infty}_{0} d{\bar p} \frac{{\bar p}^{1-\delta}}{{\bar p}^2+a^2 +4\pi^2 n^2}  \nonumber \\
    &\approx& -\frac{2 \pi^2 n^2}{\delta}+\(2\pi^2 n^2+\frac{a^2}{4}\)\ln{\left(a^2 +4\pi^2 n^2\right)}.
\end{eqnarray}
Similarly for the integral $K$, we obtain
\begin{eqnarray}
    K&=& \widetilde{K} + \int d{\bar p} \frac{{\bar p}^{3-2\delta}}{{\bar p}^2 +4\pi^2 n^2}   \nonumber\\
    &=&\widetilde{K} -\frac{2\pi^2 n^2}{\delta}+2\pi^2 n^2 \ln{\left(4\pi^2 n^2\right)}.
\end{eqnarray}
Replacing above expressions in (\ref{eB5})
We note that the divergent terms of the integrals $J$ and $K$ cancel each other out, allowing us to perform the sums in (\ref{eB5}), and  then we find the following,
\begin{eqnarray}
I(a)=I(0)+\delta I- \widetilde{K}+ \sum_{n=-\infty}^\infty\left[
\frac{a^2}{4}\ln{\left(a^2 +4\pi^2 n^2\right)} +4\pi^2 n^2 
\ln{\left(1+\frac{a^2}{4\pi^2 n^2}\right)}\right].
\label{I1}
\end{eqnarray}
The first sum above  is computed using the identity
\begin{equation}
\prod_{n=1}^\infty \left(1+\frac{a^2}{4\pi^2 n^2}\right)=\frac{2}{a}\sinh(a/2)   , 
\end{equation}
and the last one can be calculated for small $a$, using Taylor expansion, obtaining
\br
\sum_{n=1}^\infty n^2
\ln\left(1+\frac{a^2}{4\pi^2 n^2}\right)&\approx &
\sum_{n=1}^\infty \left(\frac{a^2}{4\pi^2}- \frac{a^4}{32\pi^4n^2}+
{\cal O }(a^6)\right).
\er
In this way, we get for (\ref{I1}), using the cutoff regularization
\br
 I(a)&=&I(0)+\frac{a^2}{2}\ln \sinh(a/2)+{\cal O }(a^4)+\delta I  -\widetilde{K} +a^2\left.\left(\widetilde{C}+\ln(2\pi \Lambda)+
 \Lambda\right)\right|_{\Lambda\rightarrow\infty},\qquad \mbox{}
 \label{I2}
\er
where $\widetilde{C}$ is a divergent constant. We expect that at the limit $I(a\rightarrow0)\to I(0)$, and we find
\begin{equation}
    \delta I = \widetilde{K}-\lim_{a\rightarrow0}\bigg[a^2\left.\left(\widetilde{C}+\ln(2\pi \Lambda)+
 \Lambda\right)\right|_{\Lambda\rightarrow\infty}\bigg].
 \label{b37}
\end{equation}
However, since we assume that the $I(a)$ function is smooth and continuous, we can generalize eq. (\ref{b37}) for all allowed values of $a$, i.e., those that make sense in an expansion of $a$. Thus, we have 
\begin{equation}
    I(a)=I(0)+\frac{a^2}{2}\ln \sinh(a/2)+{\cal O }(a^4).
\end{equation}
\begin{figure*}[tb]
\centering
\includegraphics[width=\textwidth]{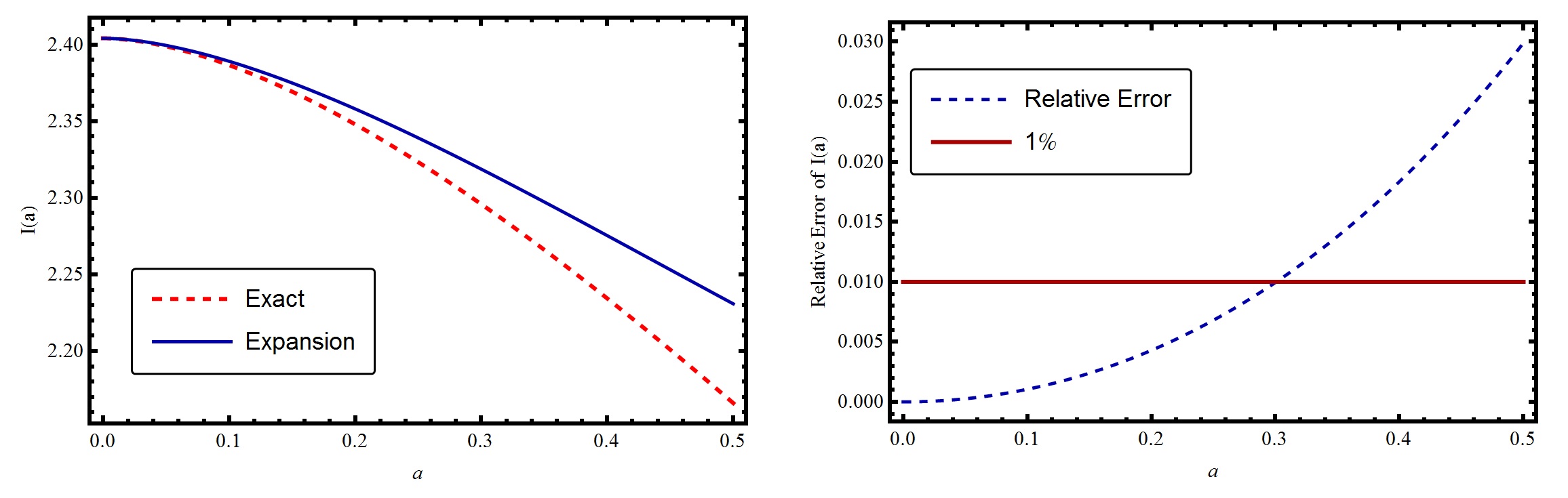}
\caption{On the left, we have plotted the values of the integral both exact numerical calculation (red dashed line) and the approximation (blue continuous line). On the right,  we have plotted the relative error of the approximated value of the integral with respect to numerical exact value. The horizontal continuous (red) line corresponds to an error of 1\%.} 
\label{integral}
\end{figure*}
\noindent In figure \ref{integral}, we have plotted the values of the integral both exact numerical calculation, and approximated calculation, and the corresponding relative error. We can notice that the relative error will be less than $1\%$ when compared with the numerical value, for $\beta m \lesssim 0.3$.   From these numerical support, we can ensure that the approximated value for the mean number of particles is reliable.

Finally, by substituting our approximation in (\ref{eB3}), we can write  down the mean number of particles formula for the massive case in the following form,

\br
   N&=&\frac{ V |\det M| T^3 }{\pi^2 |\det A|^{1/2}} \bigg(\!  \zeta(3) +\frac{m^2 T^{-2}}{4(1+\kappa^{00})}
\ln\sinh\bigg(   \frac{m}{2\sqrt{(1+\kappa^{00})}T}\bigg)+{\cal O}(( m/T)^4)
\bigg).\qquad \mbox{}
\er

%%%%%%%%%%%%%%%%%%%%%%%%%%%%%%%%%%%%%%%%%%%%%%%%%%

%%%%%%%%%%%%%%%%%%%%%%%%%%%%%%%%%%%%%%%%%%%%%%%%%%
%
\section{First correction for the massive interacting partition function}
\label{appC}
In this appendix we compute the integral in (\ref{e69}) for the massive case. 
Using (\ref{e41}) and the change of variable (\ref{novop}), we have
\begin{eqnarray}
  \int\!\! \frac{d^3{\bf p}}{(2\pi)^3}
  \frac{1}
  {[(1+\kappa^{00})p_0+\kappa^{0j}p_j](e^{\beta p_0}-1)}=\frac{|\det M| \beta^{-2}}{8\pi^3|\det A|^{1/2}(1+\kappa^{00})}\tilde{I}(a),
\label{ac1}
\end{eqnarray}
where 
\begin{equation}
\tilde{I}(a)=\int d^3\bar{\bf p} \frac{1}
{\sqrt{\bar{p}^2+a^2}[\exp(\sqrt{\bar{p}^2+a^2}-{\bf b}.\bar{\bf p})-1]},
\label{ac2}
\end{equation}
where ${\bf b}$ and  $a^2$  are given respectively  by Eqs. (\ref{e43a})  and
(\ref{eb3a}). Integral (\ref{ac2}) can not be computed exactly, and then here we will an approximated expression valid for sufficiently small $a$. For
this purpose, we first integrate in spherical coordinates, taking the polar axis along the vector
${\bf b}$. In this way we get,
\begin{eqnarray}
\tilde{I}(a)&=&2\pi \int_0^\pi \!\! d\theta \! \int_0^\infty\!\! d\bar{p} 
\frac{\sin(\theta)\bar{p}^2(\bar{p}^2+a^2)^{-1/2}}
{\exp(\sqrt{\bar{p}^2+a^2}- \cos(\theta)b\bar{ p})-1}\nonumber\\
&=&2\pi \!\int_{-1}^1 \!\!d\tau \! \int_0^\infty\!\! d\bar{p} 
\frac{\bar{p}^2(\bar{p}^2+a^2)^{-1/2}}
{\exp(\sqrt{\bar{p}^2+a^2}- b\bar{ p}\tau)-1},
\label{ac3}
\end{eqnarray}
where  we have used the change of variable $\tau=\cos(\theta)$. Using
the identity (\ref{adb1}) in (\ref{ac3}), we get 
\begin{eqnarray}
\tilde{I}(a)&=&\!-2\pi\!\int_0^\infty \!\!\!d \bar{p} \frac{\bar{p}^{2-\epsilon}}{\sqrt{\bar{p}^2+a^2}}
\!+2\pi\!\!\!\sum_{n=-\infty}^ \infty\! \int_{-1}^1 \!\!d\tau\!\! 
 \int_0^\infty \!\!\!\!d \bar{p\,}  
\frac{\bar{p}^{2-\epsilon}(\bar{p}+a^2)^{-1/2}(\sqrt{\bar{p}^2+a^2}-b\bar{p} \tau)}
{(\sqrt{\bar{p}^2+a^2}-b\bar{p} \tau)^2+4\pi^2 n^2}+\delta \tilde{I}, \nonumber \\\quad \mbox{}
\label{ac4}
\end{eqnarray}
where a regulator $(\bar{p})^{-\epsilon}$ is introduced to give sense to the change of  the order  between the sum and the integral. This change  also introduces an ambiguity, that we denote as $\delta \tilde{ I}$, and will be fixed with the known value of $\tilde I$ at $a=0$. 
For this value,  we get easily from (\ref{ac3})
\begin{equation}
\tilde{I}(0)=\frac{2\pi^3}{3(1-b^2)}.
\label{ac5}
\end{equation}
Now, for sufficiently small $a$, after expansion of the square root in (\ref{ac4}), we get
\begin{equation}
\tilde{I}(a)=I_1+I_2+I_3+\delta \tilde{I},
\label{ac6}
\end{equation}
where
\begin{eqnarray}
I_1&=&-2\pi\int_0^\infty \!\!d \bar{p} \frac{\bar{p}^{2-\epsilon}}{\sqrt{\bar{p}^2+a^2}},
\label{ac7}\\
I_2&=&\!2\pi \!\!
\sum_{n=-\infty}^\infty \int_{-1}^1 \!\!\!d\tau (1-b\tau)
\int_0^\infty \!\!\!d \bar{p} \frac{\bar{p}^{2-\epsilon}}{(1-b\tau)^2\bar{p}^2+(1-b\tau)a^2+4\pi^2 n^2},
\label{ac8}
\end{eqnarray}
and
\begin{eqnarray}
I_3&=&\pi b a^2 \sum_{n=-\infty}^\infty \int_{-1}^1 \!d\tau~\! \tau \int_0^\infty \!\!\!d\bar{p}\frac{\bar{p}^{-\epsilon}}{(1-b\tau)^2\bar{p}^2+(1-b\tau)a^2+4\pi^2 n^2}.
\label{ac9}
\end{eqnarray}
The integral (\ref{ac7}) can be evaluated in terms of the Beta function, namely
\begin{eqnarray}
I_1&=&-2\pi a^{2-\epsilon} B\(\frac{3}{2}-\frac{\epsilon}{2 },-1+\frac{\epsilon}{2}\)
\nonumber\\
&=& \frac{\pi a^2}{\epsilon}-\frac{\pi a^2}{2}\Big(1+2\ln(a/2)\Big).
\label{ac10}
\end{eqnarray}
The integrals (\ref{ac8}) and (\ref{ac9}) are evaluated using an appropriated change of variable
and 
\begin{equation}
\int_0^\infty dx \frac{x^{-\alpha}}{x^2+1}=\frac{\pi}{2\cos(\pi\alpha/2)}.
\label{ac11}
\end{equation}
In this way get respectively 
\begin{equation}
I_2=-\pi^2 \sum_{n=-\infty}^\infty\int_{-1}^1 d\tau
\frac{[(1-b\tau)a^2+4\pi^2n^2]^{(1-\epsilon)/2}}
{(1-b\tau)^{2-\epsilon} \cos(\pi\epsilon/2)},
\label{ac12}
\end{equation}
and 
\begin{equation}
I_3=\frac{\pi^2a^2b}{2} \!\sum_{n=-\infty}^\infty\!\int_{-1}^1 \!\!d\tau 
\frac{\tau [(1-b\tau)a^2+4\pi^2n^2]^{-(1+\epsilon)/2}}
{(1-b\tau)^{1-\epsilon}\cos(\pi\epsilon/2)}.
\label{ac13}
\end{equation}
Now, taking the limit $\epsilon\to 0$, separating the term $n=0$,
and  expanding  in powers of $a^2$, we obtain for (\ref{ac12})
\begin{eqnarray}
I_2=-\pi^2 a\int_{-1}^1 d\tau \frac{1}{(1-b\tau)^{3/2}}
-8\pi^3\frac{\zeta(-1)}{1-b^2}-\frac{\pi a^2}{2}\left(\frac{1}{\epsilon}+\gamma-\ln2\pi\right)
\int_{-1}^1d\tau\frac{1}{1-b\tau}
+...,\quad \mbox{}
\label{ac14}
\end{eqnarray}
where $\gamma$ is the Euler-Mascheroni constant. In a similar way,  we get for (\ref{ac14})
\begin{eqnarray}
I_3&=&\frac{\pi^2 ab}{2}\int_{-1}^1 d\tau \frac{\tau}{(1-b\tau)^{3/2}}+\frac{\pi a^2}{2}\left(\frac{1}{\epsilon}+\gamma-\ln 2\pi\right)
\int_{-1}^1 d\tau\frac{b\tau}{1-b\tau}+...    .
\label{ac15}
\end{eqnarray}
Replacing (\ref{ac10}), (\ref{ac14}) and (\ref{ac15}) in (\ref{ac6}), we see that the divergent terms at $\epsilon\to 0$  cancel out, and we obtain
\begin{eqnarray}
\tilde{I}(a)&=&\frac{\pi^2a}{2}\int_{-1}^1 d\tau 
\frac{b\tau-2}{(1-b\tau)^{3/2}}
-\frac{8\pi^3\zeta(-1)}{1-b^2}
-\frac{\pi a^2}{2}\left(1 +2\gamma +2\ln\frac{a}{4\pi}\right)+\delta \tilde{I}.
\label{ac16}
\end{eqnarray}
Now we can fix $\delta \tilde{I}$. Taking $a\to 0$ in expression above we find
\begin{equation}
\delta \tilde{I}=\frac{8\pi^3\zeta(-1)}{1-b^2}+\tilde{I}(0).
\label{ac17}
\end{equation}
Using the above result, and performing the integration in $\tau$, we  obtain
\begin{eqnarray}
\tilde{I}(a)&=&\frac{2\pi^3}{3(1-b^2)}
-\pi^2 a\left(\frac{1}{\sqrt{1-b}}+\frac{1}{\sqrt{1+b}}\right)-\frac{\pi a^2}{2}\left(1+2\gamma +2\ln\frac{a}{4\pi}\right)+...   .
\label{ac18}
\end{eqnarray}
Finally, by using eqs.  (\ref{ac18}) and (\ref{ac1}) in  eq. (\ref{e69}), we get the expression (\ref{e76}) for the partition function.

%%%%%%%%%%%%%%%  REFERENCIAS %%%%%%%%%%%%%%%%%%%%%%%%%%%%%%%%%
%\newpage

\end{document}